\documentclass[a4paper,11pt]{article}
\pdfoutput=1
\usepackage[dvipsnames]{xcolor}
\usepackage{jheppub}
\usepackage{amsfonts}
\usepackage{amsmath}
\usepackage{mathtools}
\usepackage{array}
\usepackage[markup=nocolor]{changes}
\usepackage{graphicx}
\usepackage{hyperref}
\usepackage{multirow}
\usepackage{url}
\usepackage{sectsty}

\usepackage{pgfplots}
\usepgfplotslibrary{fillbetween}
	
\usepackage{xspace}
\usepackage[utf8]{inputenc}
\usepackage[T1]{fontenc}
\usepackage{soul}
\usepackage{cancel}
\usepackage{physics}
\usepackage{slashed}
\usepackage{tabu}
\usepackage{appendix}

\usepackage{natbib}
\def\be{\begin{equation}}
\def\ee{\end{equation}}
\def\bestar{\begin{equation*}}
\def\eestar{\end{equation*}}
\def\ba{\begin{aligned}}
	\def\ea{\end{aligned}}
\def\bea{\begin{eqnarray}}
\def\eea{\end{eqnarray}}

\hypersetup{pdfstartview=FitV,colorlinks=true,linkcolor=PineGreen,citecolor=red,filecolor=black,urlcolor=PineGreen}
\sectionfont{\color{PineGreen}}
\subsectionfont{\color{PineGreen}}
\subsubsectionfont{\color{PineGreen}}

\title{Worldline theories with towers of internal states}
\author[a,b]{Steven Abel}
\author[b]{{\it and} ~Daniel Lewis}

\affiliation[a]{Institute for Particle Physics Phenomenology, Durham University, South Road, Durham, U.K.}
\affiliation[b]{Department of Mathematical Sciences, Durham University, South Road, Durham, U.K.}

\emailAdd{s.a.abel@durham.ac.uk}
\emailAdd{daniel.lewis@durham.ac.uk}

\abstract{We study particle theories that have a tower of worldline internal degrees of freedom. 
Such a theory can arise when the worldsheet of closed strings is dimensionally reduced to a worldline, in which case the tower is infinite with regularly spaced masses. But our discussion is significantly more general than this, and there  is scope to consider all kinds of internal degrees of freedom carried by the propagating particle. For example it is possible to consider towers corresponding to other geometries, or  towers with no obvious geometric interpretation that still yield a modular invariant theory. Truncated towers generate non-local particle theories that share with string theory the property of having a Gross-Mende-like saddle point in their amplitudes. This provides a novel  framework for constructing exotic theories which may have desirable properties such as finiteness and modular invariance.
 }


\begin{document} 
	\maketitle
	\flushbottom
	
	
\section{Introduction}
	This paper considers a particular modification one can make to a particle theory to make it nonlocal.	
The specific modification is to augment the particle theory expressed in the worldline formalism  \cite{feyn, Affleck:1981bma, Bern:1991aq, Strassler:1992zr,Schmidt:1993rk,Schmidt:1994zj,Bastianelli:2007pv,Edwards:2019eby}  with a tower of states that have worldline masses. If the tower is infinite it mimics the internal modes of the string, and indeed it can be directly derived by a dimensional reduction of the string to the worldline. Nevertheless one can work entirely in the worldline formalism with such a theory and still recover all the familiar properties of string physics, including its UV finiteness and even the modular invariance of its one-loop diagrams, without ever considering the geometry. Taking the worldline spectrum as the guiding principle lends a  different perspective to the finiteness of string theory. What propagates in this picture is literally a particle with an infinite number of internal harmonic oscillator degrees of freedom. In addition the spectrum of the worldline states is a more general  starting point than any particular type of string, and it leads  naturally to other physical situations, such as strings in non-trivial backgrounds, and could be of relevance for discussing higher dimensional objects and more general kinds of non-local theory. For example even a truncated tower yields stringy behaviour and yet it is a genuine particle theory, albeit a non-local one. Moreover we find other modular invariant partition functions (realisable as Borcherds products), whose {\it only} physical interpretation appears to be a that of a particle with a very specific tower of internal degrees of freedom.
	
	The original motivation for this study was to better understand the way that string theory achieves its ultra-violet (UV) completion, and its relation to non-local particle theories. Indeed, being a theory of extended objects, it is often said to be both local and non-local at the same time \cite{Eliezer:1989cr,Chaichian:1992hq,Calcagni:2007ef,Calcagni:2013eua}. For example string theory has the characteristics of a local theory in the sense that one can define a spacetime propagator in terms of a sum over conventional local particle propagators, typically an infinite sum of string excitations, Kaluza-Klein modes,  winding modes and so forth \cite{Cohen:1985sm}. Such a system would naturally be expected to have a spectral function for example. On the other hand, it is non-local in the sense that in the high momentum limit the terms from the individual modes are not physically accessible. In order to produce a single heavy resonance one might try to devise an experiment that sends a large momentum down the appropriate propagator. But heavy modes can decay to all the lighter modes below them, so their resonance peaks become broad and start to overlap. The	resulting amplitude is a smeared average over the infinite number of terms in the sum. This average, which is increasingly softened at large momentum transfers, can resemble a non-local field theory in certain kinematic regimes (indeed it has been argued that it {\it must} do so for the theory to remain unitary \cite{Chaichian:1992hq}). The expectation is that the generic softening behaviour of this non-local particle theory will for example alleviate the UV divergences in the same way as the string theory to which it corresponds. Despite any misgivings that one may have about non-local theories in general, these ones would be of special interest.
	
	As mentioned, it is natural to adopt a worldline approach for this \cite{feyn,Affleck:1981bma, Bern:1991aq, Strassler:1992zr,Schmidt:1993rk,Schmidt:1994zj,Bastianelli:2007pv,Edwards:2019eby}, in which string theory can emerge as a particle-theory-plus-corrections, corresponding to a ``projection'' of the string worldsheet to the worldline. By this method, which we shall discuss in the next section, one can arrive at a particle theory that secretly encodes all the higher dimensional properties of string theory.  Or one can truncate the worldline spectrum so as to give a non-local particle theory that mimics the first traces of stringy corrections. We will see that in a typical scattering process it yields the saddle-point behaviour of amplitudes characteristic of string theory in certain kinematic regimes (which a typical non-local theory defined in terms of infinite derivative field theories do not easily reproduce).
	
	The truncated theory also has other string-like properties. In particular, the internal degrees of freedom represented by the tower, via the worldline Green function, are 	the source of all its non-locality. In such theories the notion of ``short-distance''  must be defined with respect to the length of the worldline, and it can have only relative meaning: probing short distance means allowing vertices to coalesce relative to the diagram as a whole. This in turn means that interactions are a pre-requisite for the detection of any deviation from the standard spacetime propagator. We can conclude that the propagators of these theories show non-locality, but only at the quantum level: at tree-level they are indistinguishable from those of a local theory. This behaviour is similar to the non-local behaviour that first inspired the ``infinite-derivative-field-theory'' models of  \cite{Siegel:2003vt,Biswas:2005qr,Biswas:2011ar,Buoninfante:2018mre}.
	
	The plan of this paper is as follows. The aim of  section \ref{sec2} is to give a worldline perspective on string theory, in particular to show that  much of string physics, in particular its finiteness, can be 
	recovered if we know the form of its worldline spectrum. To this end we first study the dimensional reduction of the string worldsheet action onto a worldline. In passing, in section \ref{sec2.2}, we discuss the relation to certain particle theories with a non-local worldline theory (equivalently with a worldline theory that has infinite derivatives) that have been previously considered. Then in sections \ref{sec2.3} and \ref{sec2.4} we show how a modular invariant partition function emerges automatically from an un-truncated linear worldline spectrum (independently of any direct discussion about modular invariance). We then focus on the Green functions, which are the main ingredient in the softening of the genus one string amplitudes. In section \ref{sec3}, we use these results to consider the 
behaviour of various truncated ``modified worldline'' theories, which retain some of the crucial characteristics of string theory. In section \ref{sec4}, we consider the softening behaviour of these modified worldline theories in the hard scattering (Gross-Mende) limit. Finally, in section \ref{sec5} we discuss deformations of the spectrum that one might consider to search for other theories: some inevitably describe string theory in other geometric situations, for example orbifolds or strings in plane wave backgrounds, while others would correspond to higher dimensional objects
that do not seem to have a stringy equivalent, and to worldline systems that are modular invariant and related to Borcherds products: the latter do not have a clear geometric interpretation.

	\section{A worldline perspective on string theory}
	\label{sec2}
	We begin by elucidating the relation between non-local particle theories and string theories. In particular we would like to demonstrate that  string theory  can be understood as a simple worldline theory from which all the stringy physics including UV finiteness is seen to ``emerge''. From this point-of-view string theory can be interpreted as 
	just another non-local particle theory, but one in which its worldline spectrum happens to have a particular geometric description. In order to do this we will in the next subsection first project down string theory onto the worldline. Having done that it will then be interesting to work in the reverse direction to see how string theory can be deduced from the augmented particle theory. 
	
	\subsection{String theory as Kaluza-Klein on a Cylinder}
	\label{sec2.1}
	The traditional worldline formalism of quantum field theory treats a quantum field theory (QFT) perturbatively as a one-dimensional gravitational theory, whose universes correspond to legs of a corresponding spacetime Feynman diagram. The simplest such theory, which describes the propagation of space-time scalars, is
	\begin{equation}\label{eqn:ActionWorldlineBoson}
	\begin{aligned}
	S[X_m,g] &=  \frac{1}{4\pi \ell_s^2}\int_M d\tau  \sqrt{g} \bigg\{ g^{\tau\tau} G_{\mu\nu}\partial_{\tau}X_0^\mu \partial_\tau X_0^\nu \bigg\}\;,
	\end{aligned}
	\end{equation}
	where $X_0^\mu$ is a worldline scalar field, interpreted as the embedding of a bosonic particle in $D$-dimensional spacetime, $G_{\mu\nu}$ is some background metric and $\ell_s$ is some length scale. The central theme of this paper is that a great deal of stringy physics and much else besides can be understood by the addition of a large tower of Kaluza-Klein (KK) particles to this theory: namely the action becomes
	\begin{equation}\label{eqn:ActionGaugeFixed1}
	\begin{aligned}
	S[X_m,g] &~=~  \frac{1}{2\pi \ell_s^2}\int_0^1 d\tau  \sqrt{g} \bigg\{ \frac12 g^{\tau\tau}|\partial_{\tau}X_0|^2+\sum_{m=1}^N \bigg(  g^{\tau\tau}|\partial_\tau X_m|^2+  (2\pi)^2  f_m  |X_m|^2\bigg)\bigg\}\;,
	\end{aligned}
	\end{equation}
	where the $X_m$ are complex worldline scalar fields, $N$ is either a large integer or formally infinite (as it is in string theory), and the worldline mass-squared $f_m$ is some function of the integers. Note that for higher dimensional objects this may more naturally be a multiple-summation, so this expression is somewhat schematic. 
		
	This action is natural from the perspective of string theory, in which $f_m=m^2$. Indeed, it can be derived from a straightforward dimensional reduction of the Polyakov action for closed strings. We start from the simplest Polyakov action:
	\begin{equation}
	S[X,g] ~=~ \frac{1}{4\pi \ell_s^2}\int_{\Sigma} d^2 \sigma \sqrt{g} \bigg\{ g^{ab}G_{\mu\nu}\partial_a X^\mu \partial_b X^\nu \bigg\}\;.
\end{equation}	
As usual, we could add a cosmological constant and an antisymmetric $B_{\mu\nu}$ field, but we will not for this specific study (although we will make some general comments later). Let us summarise the other assumptions and conventions before proceeding: throughout  we will assume the target space metric $G_{\mu\nu}$ is flat and can be Wick rotated to Euclidean signature. The path integral involves summing over all Riemann surfaces $\Sigma$ and this topological dependence has been notationally suppressed in $S[g,X]$. We extend this sum to include Riemann surfaces with boundaries - so that in a moment we will be taking $\Sigma$ to be a cylinder. The manifold $\Sigma$ is parametrised, at least locally, by $\sigma_{a}$ with $a=1,2$, and has a Euclidean metric (so that the Minkowski time-like coordinate is $\sigma_0 = i\sigma_2$). Assuming that there are $D=26$ $X^\mu$ fields, the theory has no conformal anomaly and so the conformal symmetry of the action is a true gauge symmetry. In the following we often drop the $\mu$ subscripts and  take just one $X$ field, but this is merely for notational simplicity. For the previous statements to hold without including ghost fields, we assume light-cone gauge when necessary.

Now let us make a dimensional reduction of this theory, considering as mentioned the cylinder. This is simply the propagator in string theory, or in other words, the amplitude for a string to move between two given circles in space-time. To parameterise it we let $\sigma_2 \in [0,1]$ and take a periodic $\sigma_1\in [0,1]$. Diffeomorphisms and Weyl invariance can be used to completely gauge fix the cylinder metric to $ds^2 = d\sigma_1^2 + T^2d\sigma_2^2$, where the parameter $T$ is the one Teichm\"uller parameter of the cylinder, analogous to the Schwinger parameter for a line segment. It will be more convenient, however, to retain some gauge freedom. Indeed, working on the covering space $\mathbb{C}$ parametrised by $z=\sigma_1+ iT\sigma_2$, the diffeomorphism $z\mapsto z+z_2 A$ twists one of the boundary circles in $\partial \Sigma$ by an amount $A$. Clearly we should identify $A \sim A+1$ and regard $A$ as being a one-dimensional gauge field. The corresponding metric is $g = d\sigma_1^2 + 2Ad\sigma_1\otimes d\sigma_2 + (T^2 +A^2)d\sigma_2^2$. We will say that the cylinder is untwisted when $A=0$. Of course, this metric also descends to that of a torus (when we also identify $z \sim z+iT$ in the covering space), in which case $A$ is precisely the real part of the modular parameter, usually denoted $\tau_1$. 

We then carry out a Kaluza-Klein compactification on the cylindrical worldsheet, with the ansatz 
\begin{equation}
	X(\sigma_1,\sigma_2) ~=~ \sum_{m\in\mathbb{Z}}X_m(\sigma_2)e^{2\pi i m \sigma_1}~,
\end{equation}
together with a reality condition $X_{-m} = X_m^\dagger$, and with metric\footnote{In principle, we could consider the more general ansatz where the written metric is multiplied by $e^{2\phi}$. This conformal factor is cancelled in the action and so is irrelevant. The field $\phi$ does turn up in the dimensional reduction of the Ricci scalar, where is has the interpretation of being an emergent spatial coordinate, belonging to a dimensionally reduced Liouville action.} (which depends only on $\sigma_2$)
\begin{equation}
	g_{ab}~=~\begin{pmatrix} 1 & A_\tau \\ A_\tau & g_{\tau \tau}+A_\tau^2 \end{pmatrix}\;.
\end{equation}
In the above, we are anticipating the notation $\tau\coloneqq \sigma_2$ as the Euclidean worldline coordinate, in which case the gauge field is $A=A_\tau d\tau$. We have also left $g_{\tau \tau}$ as a general one-dimensional metric (as opposed to fixing it to $T^2$ as in the previous paragraph). Obviously, the $X_m$ are then simply related to the string oscillator modes. Upon dimensional reduction and retaining all the fields, the resulting worldline action is
\begin{equation}\label{eqn:Action}
\begin{aligned}
	S[X_m,A,g] &~=~  \frac{1}{2\pi \ell_s^2}\int_0^1 d\tau \sqrt{g_{\tau\tau}} \bigg\{ \frac12 g^{\tau\tau}|D_{\tau}X_0|^2+\sum_{m=1}^\infty \bigg( g^{\tau\tau} |D_\tau X_m|^2+ (2\pi m)^2 |X_m|^2\bigg)\bigg\}\;,
\end{aligned}
\end{equation}
where the covariant derivative is $D_{\tau} X_m= (\partial_{\tau}-2\pi im A_{\tau})X_m$. 
As discussed, because it is pure gauge, we can locally set $A=0$ and enforce the associated Gauss constraint $\sum_m \text{Im}(X^\dagger_m \partial X_m)=0$, which is equivalent to the level-matching conditions. The gauge fixed action with $A=0$ and $g_{\tau\tau}=T^2$ becomes
\begin{equation}\label{eqn:ActionGaugeFixed}
\begin{aligned}
	S[X_m,T] &~=~  \frac{1}{2\pi \ell_s^2}\int_0^1 d\tau  \bigg\{ \frac{1}{2T} (\partial_{\tau}X_0)^2+\sum_{m=1}^\infty \bigg( \frac{1}{T} |\partial_\tau X_m|^2+ T (2\pi m)^2 |X_m|^2\bigg)\bigg\}\;.
\end{aligned}
\end{equation}
Note that the $U(1)$ gauge symmetry described by $A$ acts on the $X_m$ fields \textit{simultaneously}: under $A\mapsto A+d\alpha$ we have $X_m \mapsto  e^{2\pi i m \alpha}X_m$ for all $m$, so it applies a single constraint to the spectrum. Equivalently, since each $X_m$ has a standard global $U(1)$ symmetry, the gauge symmetry is taken to be the diagonal of the product ``$U(1)^\infty $'' global symmetry. Although locally $A$ can be gauged away, it will become important when the worldline topology is non-trivial. 

As anticipated this dimensional reduction gives us a concrete way to view the string as an infinite tower of complex one dimensional fields with worldline masses. Note that the field $X_0$, the centre of mass of the string, is to be interpreted as the worldline coordinate field, whereas the massive fields are harmonics corresponding to fluctuations of the extended string around it. However none of these elements is unusual from the perspective of the particle theory. In the worldline formalism internal degrees of freedom are {\it always} realised by adding extra worldline fields, and constraints on the spectrum as worldline gauge symmetries. The main difference here is that the tower of states that augments the particle theory is infinite (hence they describe a continuous object) and they are massive (hence their effect is non-local in space-time).

	\subsubsection*{Relation to Theories with {\it Worldline} Non-Locality} 
	It is an interesting fact that the dimensionally reduced action above, say with $A=0$, corresponds precisely to a worldline theory that has non-locality on the worldline, and which involves only a single worldline field. Equivalently 
	it can be written as a worldline field theory that contains an infinite number of derivatives. To make the connection we consider actions with worldline nonlocality of the form   \begin{equation}\label{eqn:NonLocalWorldLine}
	S ~=~ \frac{1}{2\pi \ell_s^2} \int d\tau \sqrt{g} \; G_{\mu\nu}\;  g^{\tau\tau}\partial_{\tau} {X}^\mu(\tau) F(\sqrt{\Box})\partial_\tau X^\nu(\tau)\;,
\end{equation}
where we define $\sqrt{\Box} = \sqrt{g^{\tau\tau}\partial_\tau^2}$, which in terms of the einbein $e(\tau)$ can be written $\sqrt{\Box} = e^{-1}\partial_\tau$. Gauge fixing $g_{\tau\tau}=T^2$ and making the particular choice 
\begin{equation}\label{eqn:TanhChoice}
	F(\sqrt{\Box}) ~=~ \frac{\tan(\frac12 \sqrt{\Box})}{\sqrt{\Box}}\;,
\end{equation}
gives the worldline theory that was considered in  \cite{Kato:1990bd}. (Further aspects\footnote{Note that we are working in Euclidean signature on the worldline, whereas these papers work in Minkowski.} of this non-local action have been explored in \cite{Kato:1990bd,Cheng:2008qz,Calcagni:2013eua}.)
This theory can be seen to be equivalent to \eqref{eqn:ActionGaugeFixed} through the identity
\begin{equation}
	\frac{1}{2z\tan(z/2)} ~=~ \frac{1}{z^2}+ 2\sum_{n=1}^\infty \frac{1}{z^2-(2\pi n)^2}\;.
\end{equation}
That is, the formal inverse of the Green function of $\tan(\frac{1}{2T}\partial_\tau)\cdot \frac{2}{T}\partial_\tau$ is the infinite sum of individual Green functions $G_m=1/(-\partial_\tau^2 + (2\pi m T)^2)$. It is then reasonable to suppose that the nonlocal action in \eqref{eqn:NonLocalWorldLine} is equivalent to an infinite number of fields having action involving kinetic term $-\partial_\tau^2+(2\pi m T)^2$, with the two being related to each other  by integration by parts. 

An alternative approach to the one we are pursuing here,  would be  instead to consider all nonlocal worldline actions as the starting point, rather than 
theories augmented by towers of massive worldline states. We have favoured this latter approach because it connects directly to the ultimate objects such as partition functions and amplitudes that one would like to calculate. Moreover the two options are qualitatively different: a nonlocal worldline theory implies an infinite number of derivatives which in turn implies an infinite tower of additional wordlline states. By contrast, and as we shall do later, it is interesting to truncate the tower of states to find a genuine particle approximation (albeit a nonlocal one) to the string theory.

	\subsection{String from particles: propagators}
	\label{sec2.2}

Let us now reverse direction. Suppose we were given a particle theory of the form in Eq.\eqref{eqn:ActionGaugeFixed}, and knew nothing about its geometric interpretation. Could we recover stringy behaviour from the worldline action? Consider the propagator of a point particle to travel from $x_0$ to $y_0$, which will be expressed in the form of an integral over some kernel $K(x_0,y_0;T)$, where $T$ is the Schwinger parameter (which for us is always in string units). 

In the worldline description we should treat this as a tower of particle propagators with worldline masses, travelling along a line segment with Schwinger parameter $T$. Consider the action corresponding to a single massive field $X_m$, gauge fixing $A=0$:
\begin{equation}
\label{eq:Sm}
\begin{aligned}
	S_m[X_m,T] &~=~  \frac{1}{2\pi \ell_s^2}\int_0^1 d\tau  \bigg\{ \frac{1}{T} |\partial_\tau X_m|^2+ T (2\pi m)^2 |X_m|^2\bigg\}\;.
\end{aligned}
\end{equation}
The amplitude for a single massive state $X_m$ to go from initial boundary condition $x_m$ to final $y_m$ is given by the Mehler kernel:
\begin{equation}
	\begin{aligned}K_m(x_m,y_m;T) &~=~  \int_{X_m(0)=x_m}^{X_m(1)=y_m}  \mathcal{D}X_m e^{-S[X_m,T]}\\
	&~=~  \frac{\pi}{2} \cdot \frac{m}{\sinh(2\pi mT)} e^{-\frac{m((|x_m|^2+|y_m|^2)\cosh(2\pi mT) - 2\text{Re}(x_m \cdot y_m))}{\ell_s^2 \sinh(2\pi mT)}}\;,
	\end{aligned}
\end{equation}
which can be found by the usual heat kernel methods. The expression in the exponent is the on-shell action, whilst the prefactor arises from a regularization of a functional determinant. It is important to note that this regularization is physically meaningful in that it gives rise to a counterterm in the vacuum energy. For $D$ identical fields, the determinant (and therefore prefactor) gains a power $D$.
For $m=0$ by contrast, one simply finds the usual massless particle propagator (this also corresponds to taking $m\rightarrow 0$ in the above formula and taking the square root since $X_0$ is real), which we write for convenience:
\begin{equation}
	K_0(x_0,y_0;T)~ =~ \frac{1}{2T^{1/2}}e^{-\frac{(x_0-y_0)^2}{4\pi \ell_s^2 T}}\;.
\end{equation}
With $D-2$ such fields the denominator in the prefactor aquires an exponent $(D-2)/2$.

Putting everything together, in $D$ target-space dimensions and hence with $D-2$ independent fields (assuming light-cone gauge), the total propagator 
becomes a product over all (infinitely many) fields, each with boundary conditions $x_m^\mu$ and $y_m^\mu$. We can write this product as a regularised limit\footnote{In this expression we drop constant prefactors which are irrelevant.}:
\begin{align}
\label{eq:214}
	K(\{x^\mu_m\},\{y^\mu_m\};T) &~=~ \lim_{N\rightarrow \infty} \, \frac{e^{-\frac{(x_0-y_0)^2}{4\pi \ell_s^2 T}}}{(2T)^{{(D-2)}/2}} ~\bigg(\prod_{m=1}^N \frac{m}{\sinh(2\pi mT)}\bigg)^{D-2}  \\
	&\hspace{4cm} \times~ e^{-\sum_{m=1}^N \frac{m((|x_m|^2+|y_m|^2)\cosh(2\pi mT) - 2\text{Re}(x_m \cdot y_m))}{\ell_s^2 \sinh(2\pi mT)}} \; , \nonumber\\
	&~=~ \frac{e^{-\frac{(x_0-y_0)^2}{4\pi\ell_s^2 T}}}{(2T)^{{(D-2)}/2}} \,\frac{1}{\eta(2iT)^{2(D-2)}}~
	e^{-\sum_{m=1}^\infty \frac{m((|x_m|^2+|y_m|^2)\cosh(2\pi mT) - 2\text{Re}(x_m \, y_m))}{\ell_s^2 \sinh(2\pi mT)}}\;.\nonumber 
	\end{align}
 As a simple example, when the initial and final states are point-like, all the higher harmonics vanish and the propagator becomes 
\begin{equation}
\label{eq:kernel}
	K(x_0,x_m=0,y_0,y_m=0;T) ~=~ \frac{1}{(2T)^{(D-2)/2}|\eta(2iT)|^{2(D-2)}}~ e^{-\frac{(x_0-y_0)^2}{4\pi \ell_s^2 T}}\;.
\end{equation}

Note the appearance at this stage (already at tree-level) of the Dedekind eta function $\eta(\tau)$, which has arisen from zeta function regularization: $\prod_{m=1}^\infty \sin(2\pi m\tau) \overset{\zeta}{=} \sqrt{2}\eta(2\tau)$ for $\tau \in \mathbb{C}$.
This expression is equivalent to that of the sum of all the standard QFT propagators over the string spectrum. It is in this sense that we say (as in the introduction) that the propagator is that of a local field theory. 
This provides a very explicit version of the expressions given in \cite{Cohen:1985sm} but is also to be expected from open/closed string duality.

	\subsection{Sewing the cylinder into a torus}
		\label{sec2.3}

Let us now see how the  Schwinger $T$ parameter above gets promoted to the full complex Teichm\"uller parameter when we begin to do loop diagrams.
The simplest example is naturally that of the one-loop partition function, for which we should glue the final and initial endpoints in the above propagator. However, having previously chosen the gauge $A=0$, one should now enlarge the number of boundary conditions to take account of the ability of the  fields to go through a gauge transformation as they go around the loop. That is, we take $x_m \sim x_m e^{2\pi i m A}$  for all $m$ with arbitrary $A \in (-1/2,1/2]$. We can think of the different gauge transformations  $A$ as defining different sectors of the theory, with the path integral having to go over all sectors. By sewing together the ends of the cylinder with this twist included, the partition function after integrating over all possible regularised kernels becomes:
\begin{align}
\label{eqn:PartFctKernel}
	\mathcal{Z}(A,T) &~=~ \lim_{N\rightarrow \infty} \prod_{m=1}^N \int_{-\infty}^\infty  dx_m^\mu K(\{x^\mu_m\},\{x^\mu_me^{2\pi i m A}\};T)~,\nonumber \\
	&~=~  \lim_{N\rightarrow \infty} \prod_{m=1}^N \int_{-\infty}^\infty  dx_m^\mu \frac{1}{T^{(D-2)/2}} \bigg(\frac{m}{\sinh(2\pi mT)}\bigg)^{D-2} e^{-\frac{m|\sin(\pi m (A+iT))|^2|x_m|^2}{\sinh(2\pi mT)}}~,\nonumber \\
	&~=~ \lim_{N\rightarrow \infty} \frac{1}{T^{(D-2)/2}}  \prod_{m=1}^N \frac{1}{|\sin(\pi m (A+iT))|^{2(D-2)}} \;.
\end{align}
Upon zeta function regularisation of the product, one recovers the Dedekind eta function of the full one-loop partition function:
\begin{equation}
\mathcal{Z}(A,T) ~=~ \frac{1}{T^{(D-2)/2}|\eta(A+iT)|^{2(D-2)}}\;.
\end{equation}
Hence we naturally obtain the string torus partition function with $A+iT$ being the Teichm\"uller parameter. (Note that factor of 2 in the argument of $\eta$ compared to \eqref{eq:kernel}
is effectively the result of a double-angle identity.)

There are two remaining integrations to be done: over $T \in (0,\infty)$ and $A \in (-1/2,1/2)$. However, as is well known, there is a large degeneracy in the sewing procedure, and again we wish to put this in the worldline language.  First, a $1/T$ factor in the measure arises in the same way as it does in any worldline approach: the same circle arises from $\sim 2\pi T$ ways of sewing the dimensionally reduced line segment together (each point on the circle can be cut to give an equivalence class of line segments). Furthermore because we have assumed lightcone gauge, we should include a second $1/T$ factor coming from the two remaining zero modes belonging to the lightcone coordinates $X^\pm$ (alternatively this factor would be seen in a proper treatment of the Polyakov action involving ghost fields, these ghosts would then also appear in the dimensionally reduced worldline theory).

Next, the partition function is now exhibiting an unasked for modular $SL(2,\mathbb{Z})$ symmetry. The real part of this symmetry, i.e. $A\rightarrow A+1$, is inevitable as we put it in by hand to make the gauge symmetry compact. Meanwhile $T\rightarrow 1/T$ symmetry was already a symmetry of the kernel in Eq.\eqref{eq:kernel}, before we ever got to the one loop diagram: we should therefore not attempt to integrate over more than one representative $T$ when we sew the propagator into a loop. 
Hence the restriction to the fundamental domain of $SL(2,\mathbb{Z})$ when doing the integral over $A$ and $T$ is again something that can be required by a worldline particle theory
without ever making a reference to a worldsheet torus. Generally the worldline spectrum can be such that it ``grows'' such a geometric stringy interpretation, but this does not appear to be a necessity.
	\subsection{Green Functions and Vertex Operators}
	\label{sec2.4}
For amplitudes, we need the worldline Green functions for \eqref{eqn:Action}. We restrict ourselves to line segments and circles, although more generally, one can join these together to make worldgraphs. It is clear from the action in \eqref{eqn:Action} that
\begin{equation}
\label{eq:orth}
	\langle X_m^{\dagger}(v_1) X_n(v_2) \rangle_{A} ~=~ 2\pi \ell_s^2\; \delta_{m,n}G_{m}(v_{12};A,T)\;,
\end{equation}
where $G_m(v_{12};A,T)$ is the Green function corresponding to the kinetic operator
\begin{equation}
	L_m ~=~ -T^{-1}(\partial_v-2\pi im A)^2 + T m^2\;,
\end{equation}
omitting constant prefactors. The $A$ in the above correlation function indicates that the correlation function is to be taken in a sector with $A$ fixed. In the path integral one can either gauge fix (at least locally) to $A=0$ as usual, or one could integrate over all $A \in (-1/2,1/2]$ when, as in the partition function example, propagators are sewn together with a possible gauge transformation. Over the real line, the Green function satisfying $LG_m(v_{12})=\delta(v_{12})$ is easily found to be
\begin{equation}
	G_m(v_1,v_2;A,T) ~=~ \begin{cases} \frac{1}{4\pi |m|} e^{2\pi i m Av_{12}}e^{-2\pi |m| |v_{12}|T} & \text{ if }m\neq 0 \\
	 \frac12 T|v_{12}| & \text{ if }m= 0 \;.\end{cases}
\end{equation}
Over a circle parametrised by $v \in [0,1]$, one instead finds
\begin{equation}\label{eqn:GmCirc}
	G_m^\circ (v_1,v_2;A,T) ~=~ \begin{cases}
		\frac{1}{4\pi |m|} \sum_{k\in \mathbb{Z}}e^{2\pi i m A(k-v_{12})}e^{-2\pi |m| |k-v_{12}|T}& \text{ if }m\neq 0\\
			\frac{T}{2}(v_{12}^2 - |v_{12}|) & \text{ if }m=0 \;,\end{cases}
\end{equation}
where the latter satisfies $LG_0(v{12})=\delta(v_{12})-1$, subtracting the zero mode piece. The latter is easily obtained from the former via the method of images.

\subsubsection*{Vertex Operators} The next ingredient for constructing general graphs is the worldline vertex operator. The relevant vertex operators for the field $X_m$ are of the form $V_{m;0}=e^{ i p \cdot X_m}$, however this is not gauge invariant: under a gauge transformation   with gauge parameter $g=e^{2\pi i u}$, it transforms as $e^{ i p \cdot X_m}\mapsto e^{ i p \cdot X_m e^{2\pi i m u}}$. Thus there is a class of gauge-equivalent vertex operators $V_{u;m}=e^{ i p \cdot e^{2\pi i m u}X_m}$ with $u\in [0,1]$, and 
a gauge invariant vertex operator
\begin{align}
V~=~ \int du ~~\cdot ~~\prod_m e^{ i p \cdot e^{2\pi i m u}X_m}~
\end{align}
Note that all states are emitted at the same position on the worldline (i.e. every leg of the worldline universe has a whole tower on it) and the only conserved momentum belongs to the zero mode. For would-be gluons, the vertex operators are analogously composed out of 
\begin{align}
V_{g} ~=~ \int du ~~\cdot ~~ \sum_n \partial X_n \prod_m  e^{i p \cdot e^{2\pi i m u}X_m}~.
\end{align}

As an example, consider placing two tachyon vertex operators on a circle. Then, summing over sectors $(A)$, and also over all gauge equivalent vertex operators \textit{and} the whole tower of Kaluza-Klein modes, gives an amplitude
\begin{equation}
\begin{aligned}
	{\cal A}_2~ &=~ {\prod_{m_1,m_2\in \mathbb{Z}}}\int d\mu \int_0^1 dv\langle V_{u_1;m_1}^\dagger (p,0) V_{u_2;m_2}(q,v)\rangle_A\\
	 ~&= ~\int d\mu  \int_0^1 dv e^{-p\cdot q \sum_{m\in\mathbb{Z}}e^{2\pi i m u}2\pi\ell_s^2 G_m(v;A,T)}\mathcal{Z}(A,T)\;,
\end{aligned}
\end{equation}
where we use \eqref{eq:orth} and where $\mathcal{Z}(A,T)$ is the partition function calculated above in the sector $A$, and the integral $\int d\mu$ denotes integration over all $T$, all $A\in (-1/2,1/2)$ and $u_1,u_2\in (0,1)$.

\subsubsection*{Comparison with string theory}
Let us compare the above expressions to the more conventional expressions for a string genus one Green function. 
	Schematically string theory one-loop amplitudes are of the form
	\begin{equation}\label{eq:FourPointAmpltd}
		\mathcal{A}_n ~=~ \int_{\mathcal{F}} d^2 \tau \; \mathcal{Z}(\tau) \int d^2 z_i \prod_{\substack{i=1\\j<i}}^n e^{-k_i \cdot k_j 2\pi\ell_s^2 G(z_{ij};\tau)}\;,
	\end{equation}
	where now we integrate over the Teichm\"uller parameters of the torus $\tau=\tau_1+i\tau_2$, and the vertex operator positions are two-dimensional (complex) numbers $z_i$, integrated over the torus $T(\tau)$ with modular parameter $\tau$. The function $G(z_{ij};\tau)$ is the Green function on the torus, see equation \eqref{eq:StringyCorrectionsPropagator}. There can also be complex prefactors to the exponential which depend on the precise vertex operators - we ignore these since they are irrelevant for our discussion.
For a torus with modular parameter $\tau=\tau_1+i\tau_2$, the Green function is
	\begin{equation}
	2\pi\ell_s^2 G(z;\tau)~=~\langle X(z)X(0)\rangle_{T^{2}}~=~-\frac{\ell_s^2}{2}\bigg(\ln\left|\frac{\vartheta_{1}(z)}{\vartheta_{1}^{\prime}(0)}\right|^{2}-\frac{2\pi z_{2}^{2}}{\tau_{2}}\bigg)\;,\label{eq:GreensTheta}
	\end{equation}
	where $z=z_1+iz_2=u+\tau v$ is a coordinate on the torus and $\vartheta_{1}(z)$ is a Jacobi theta function. A Fourier expansion of $G(z)$ gives \cite{Green:1999pv}
	\begin{align}
          \label{eq:StringyCorrectionsPropagator}	
          G(z,\tau) & ~~=~~ \frac{\tau_{2}}{2} (v^{2}-|v|)\;+ \; \sum_{\substack{m\neq0\\
			k\in\mathbb{Z}}}
			\frac{1}{4\pi |m|}e^{2\pi im(u+\tau_{1}(k+v))}e^{-2\pi\tau_{2}|m||k-v|}
	 \\
	& ~~\qquad\qquad\qquad\qquad\qquad\qquad~+~2\ln2\pi\;+~2\sum_{\substack{m\neq0 \nonumber \\
			k\geq1
		}
	}\frac{1}{4\pi|m|}e^{2\pi ikm\tau_{1}}e^{-2\pi k|m|\tau_{2}}~. 
	\end{align}
	
	The field theory limit involves taking $\alpha'=\ell_s^2 \rightarrow0$ keeping $t=\pi\alpha'\tau_{2}$ fixed, which plays the role of the Schwinger parameter.
	As indicated above, we can therefore conceptually separate the torus Green function into two parts. The first term involving $v^{2}-|v|$, which we have already seen as arising from the $X_0$ Green function, is the only piece which survives in the field limit. The second term of (\ref{eq:StringyCorrectionsPropagator}) consists of  the infinite tower of stringy oscillations, which become important when $\alpha'$ is non-zero. In the following we will see that they play a central role in softening the string amplitudes from a particle point of view. There is also a function of $\tau$ only (a zero mode) on the second line which is irrelevant for our purposes.
Comparing to the  the circle Green function in \eqref{eqn:GmCirc}: above, we see that the torus Green function is a sum over the $X_m$ Green functions:
\begin{equation}
	2\pi \ell_s^2G(z=u_{12}+\tau v_{12};\tau) = \sum_{m \in \mathbb{Z}}\langle \big(e^{-2\pi i m u_1}X^\dagger_m(v_1)\big) \big(e^{2\pi i m u_2}X_m(v_2)\big)\rangle|_{T=\tau_2,A=\tau_1}\; + \; \text{zero mode}\;.
\end{equation}

	\section{Truncation:  ``mock'' string theory}
	\label{sec3}
We have seen how bosonic closed string theory defined on a cylinder reduces to a particle theory with an infinite tower of fields. However such a theory is arguably no longer a particle theory once it is regulated: it is not truly one-dimensional in the same way that any Kaluza-Klein reduced theory which retains all the harmonics is not truly lower dimensional. 

To make a genuine particle theory, one can truncate the number of fields so that the index $m$ now runs from $0$ to $N$. This `pruning' makes the theory a true theory of particles that, as we shall see, retains some physical properties of its stringy cousin. However, as in Kaluza-Klein reduction, such a truncation does not continuously approach the string theory as $N\rightarrow \infty$, in the same sense that any Kaluza-Klein theory which retains some but not all the higher harmonics is not necessarily a good approximation either. The discontinuity lies in the regularisation of the string theory, which is not possible for any finite $N$. 

	Truncating to $N$ fields leaves the action
\begin{equation}\label{eqn:TruncatedAction}
\begin{aligned}
	S[X_m,A,g] &~=~  \frac{1}{2\pi \ell_s^2}\int_0^1 d\tau \sqrt{g_{\tau\tau}} \bigg\{ \frac12 g^{\tau\tau}|D_{\tau}X^\mu_0|^2+\sum_{m=1}^N \bigg( g^{\tau\tau} |D_\tau X_m|^2+ (2\pi m)^2 |X_m|^2\bigg)\bigg\}\;,
\end{aligned}
\end{equation}
which is now well-defined on the worldline. Such a truncation is reminiscent of matrix models. The corresponding one-loop diagram is as in \eqref{eqn:PartFctKernel}, but again we do not take the $N\rightarrow \infty$ limit:
\begin{equation}\label{PartFctTruncated}
	\mathcal{Z}(A,T) = \frac{1}{T^{d/2}}\prod_{m=1}^N 4^d|q^{-m/2}|^{-2d}|1-q^{m}|^{-2d}\;,
\end{equation}
where $q=e^{2\pi i (A+iT)}$ and $d=D-2$. Of course, the Green functions of such a theory remain the same as in the previous section. With the $U(1)$ gauge symmetry, the gauge invariant vertex operators are inherited from $\int dg \; g \cdot e^{ip\cdot \sum X_m} = \int du e^{ip \cdot \sum_m e^{2\pi i m u}X_m}$ as before.

What does this theory look like in spacetime? Since there is no obvious second quantization of a theory with worldline masses, no spacetime Lagrangian seems to be available. The transition from a first quantized to second quantized theory would involve replacing fields $X^\mu _m$ by states $\ket{X^\mu _m}$. This theory then has an infinite tower of higher spin states, just as in string theory, which can indeed be read off from a $q$-expansion of the spectrum \eqref{PartFctTruncated}:
\begin{equation}
\mathcal{Z}(A,T) ~=~ \frac{1}{T^{d/2}} \cdot 4^{Nd}|q|^{N(N+1)d/2}\cdot \prod_{m=1}^N \sum_{r,s\geq 0}c_{r,s}q^{mr d}\bar{q}^{ms d}\;,
\end{equation}
for appropriate $d$-dependent degeneracies $c_{r,s}$. The imaginary part involves the unlevel matched states with $r\neq s$. Enforcing level matching shows that there is an infinite tower of states with integer mass levels starting at $N(N+1)d$. In this one-dimensional theory, there is freedom to add a cosmological constant to the action which would decrease or increase the physical mass. For example, by introducing a negative cosmological constant on the worldline one can tune the above expression so that the tower begins with massless scalars\footnote{For the transition to string theory in the formal limit $N\rightarrow \infty$, counterterms to the cosmological constant should be generated which regulate the sum ``$1+2+3+\cdots$'' to $-1/12$, thereby introducing the tachyon.}.

As mentioned, prospective UV finiteness is the missing element in a truncated theory because the truncation explicitly breaks the $T\rightarrow 1/T$ symmetry, and the integration region of $A+iT$ is no longer restricted to the usual fundamental domain of $SL(2,\mathbb{Z})$. Therefore additional regularisation is required for the truncated theories. Although they are non-local they are not UV finite. We note however that in principle an expansion of $X_m(\tau)$ into modes that are then truncated in the same way as the $m$-sum {\it would} retain $T\rightarrow 1/T$ symmetry. Presumably it would then be possible to let $N\rightarrow \infty$ continuously.

\subsection*{Discrete Gauge Symmetry}

We should briefly mention a second complementary pruning one can do to the theory which is to instead explicitly break the $U(1)$ gauge symmetry into a discrete gauge symmetry, that is only realise a gauged $\mathbb{Z}_n$ symmetry in the worldline action. Whilst this may seem un-natural from the string point of view, it will nevertheless be interesting to broaden our class of theories this way.

Beginning with the ungauged version of the action \eqref{eqn:TruncatedAction}, one could choose a $\mathbb{Z}_n$ subgroup of $U(1)$ to gauge instead of the full $U(1)$. Then the path integral over $A$, previously just an integral $\int_{-1/2}^{1/2}dA$ becomes a sum over defects $g\in\mathbb{Z}_n$ placed in the one-dimensional spacetime, which act as $g\cdot X_m = e^{2\pi i k/n}X_m$ with $g=e^{2\pi i k/n}\in \mathbb{Z}_n$. This is equivalent to discretizing the $A$ integral. The simplest case is to take $n=2$, in which case  we have two sectors, one equivalent to $A=0$ and the other equivalent to $A=1/2$. With the $\mathbb{Z}_n$ gauge symmetry, the gauge invariant vertex operators are inherited from $\sum_g g\cdot V_g = \sum_{k=0}^n e^{ip \cdot \sum_m e^{2\pi i m k/n}X_m}$.

\section{Softening of Amplitudes}
	\label{sec4}
Given the worldline theory cast in form  \eqref{eqn:TruncatedAction}, we are now able to consider the role of the exponential corrections in the particle one-loop Green function. In particular when asking what makes string theory `soft' in the UV, it is typical to invoke the fixed-angle Gross-Mende regime, where string theory famously has exponentially damped amplitudes. As we will see in this section, exponential damping in such a regime is actually provided by the first exponential correction in the Green function coming from the lowest mode in the tower. That is, even a non-local theory truncated to $N=1$ will show the same damping. Let us examine this in more detail.

\subsection{Gross-Mende Softening of Amplitudes in String Theory}\label{sec:Gross-MendeSoftening}
An ordinary scalar QFT that has been Wick rotated to Euclidean signature has a two point function which at one loop is schematically given by
\begin{equation}
	\mathcal{A}_2 ~=~ \int_0^\infty dT \mathcal{Z}(T) \int dx_1 dx_2 T^{-1} e^{-Tx_{12}(1-x_{12})p^2}\;.
\end{equation}
where $x_{12}=x_2-x_1$ and where $\mathcal{Z}(T)$ is the circle partition function\footnote{For example, with just one scalar field $\mathcal{Z}(T)=T^{-D/2}e^{-m^2 T}$.}. The function $Tx_{12}(1-x_{12})$ is of course precisely the Green function for the massless worldline scalar field on a circle of radius $T$ with vertex operator punctures at $x_1$ and $x_2$. On a fixed $T$-subspace, a saddle point argument in $x_{12}$ shows that the largest contributions occur when the vertex operators are as far away as possible - they repel each other. However, there is no saddle point in $T$, which contributes increasingly in the UV region at $T\rightarrow 0$. Thus one does not expect an exponential softening of amplitudes.
	
	In closed string theory, such an amplitude is expected to be UV-finite due to the integration region $\mathcal{F}$ excluding the dangerous UV region. But 
on the fixed $\tau$-subspace, there is again a saddle point for the vertex operator positions $z_i$. Indeed, Gross and Mende have made the analogy to electrostatics on a torus and find that even $\tau$ gains a saddle in the string theory, as well as the vertex coordinates \cite{Gross:1987kza}. 
	
	It will be convenient to review the four-point amplitude in some more detail. We use the parametrization $z_{i}=u_{i}+\tau v_{i}$,
	where $u_{i}$ and $v_{i}$ are in $[0,1]$ for each
	$i$. We have fixed $z_{4}=0$ using the one conformal Killing vector
	of the torus. It is useful to define $s_{ij}\coloneqq-(k_{i}+k_{j})^{2}$
	and we will use the convention that $s\coloneqq s_{12}$, $t\coloneqq s_{14}$
	and $u\coloneqq s_{13}$.
	
Famously, in the fixed angle limit where $t$ and $u$ scale with $s$,  the exponent in \eqref{eq:FourPointAmpltd} has a saddle when $\hat{z}_{1}=\frac{1}{2}$, $\hat{z}_{2}=\frac{\tau}{2}$ and $\hat{z}_{3}=\frac{1}{2}+\frac{\tau}{2}$ (recall we have fixed $z_4=0$). There are other saddles, but these are subdominant. It is then a straightforward exercise to find a saddle point for $\tau_{2}$; this turns out to be 
	\begin{align}
	\hat{\tau}_{2}~&=~i\frac{K(-u/s)}{K(-t/s)}\;,\nonumber \\
	&~\simeq ~ -\frac{1}{\pi}\log \left( -\frac{t}{16s} \right)~,
	\label{eq:GMapp}
	\end{align}
	where $K(z)$ is the elliptic integral of the first kind, and the approximation is in the $|t|/s\ll 1$ limit.  The softness of string amplitudes in this kinematic limit is essentially entirely due to the existence of this saddle point.

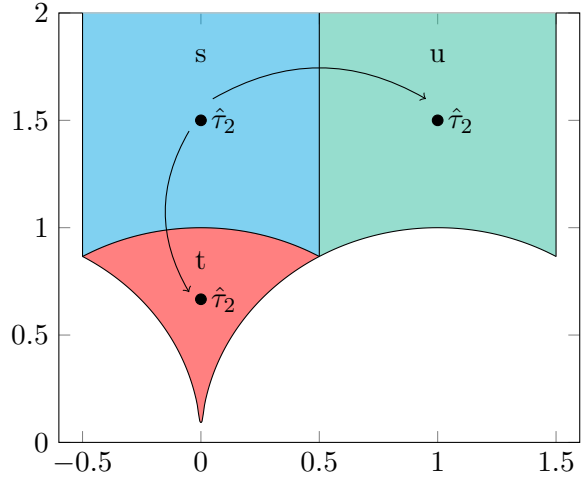
\begin{figure}\label{fig:CongSubgroup}
		\centering
		\begin{tikzpicture}
		\begin{axis}[thick,smooth,no markers,xmin=-0.6,xmax=1.6,ymin=0,ymax=2]
		\addplot[name path=A,black,domain=-0.5:0.5] {sqrt(1-x^2)};
		\addplot[name path=B,black,domain=0.5:1.5] {sqrt(1-(x-1)^2)};
		\addplot[name path=C,black,domain=-0.5:0.5,samples=100] {sqrt(2*abs(x)-x^2)};
		\addplot[name path=A2,white,domain=-0.5:0.5] {2};
		\addplot[name path=B2,white,domain=0.5:1.5] {2};
		\addplot[name path=D,black,domain=0.865:2] ({0.5},{x});
		\addplot[name path=E,black,domain=0.865:2] ({-0.5},{x});
		\addplot[name path=F,black,domain=0.865:2] ({1.5},{x});
		\addplot[Cerulean,opacity=0.5,draw] fill between[of=A and A2];
		\addplot[SeaGreen,opacity=0.5] fill between[of=B and B2];
		\addplot[red,opacity=0.5] fill between[of=C and A];
		\node at (axis cs:0,1.8) {s};
		\node at (axis cs:0,0.85) {t};
		\node at (axis cs:1,1.8) {u};
		\draw [black, fill] (axis cs:0,1.5) circle (2pt) node [right] {$\hat{\tau}_2$};
		\draw[black, fill] (axis cs:0,2/3) circle (2pt) node [right] {$\hat{\tau}_2$};
		\draw[black, fill] (axis cs:1,1.5) circle (2pt) node [right] {$\hat{\tau}_2$};
		\draw [->] (axis cs:0.05,1.6) to [out=30,in=150] (axis cs:0.95,1.6);
		\draw [->] (axis cs:-0.05,1.45) to [out=240,in=120] (axis cs:-0.05,0.7);
		\end{axis}
		\end{tikzpicture}
		\caption{Congruence subgroup $\Gamma_0(2)$, a subset of the complex plane. The Gross-Mende saddle is mapped under cycling $s$, $t$ and $u$ to copies of the $SL(2,\mathbb{Z})$ fundamental domain as shown.}
	\end{figure}

Compare this situation to its field theory limit, $\alpha'\rightarrow 0$ with fixed $t=\pi\alpha'\tau_{2}$. First, we break the string amplitude into three pieces
	\begin{align}\label{eqn:stuString}
	\mathcal{A}_{4}(s,t,u)~&=~2\mathcal{A}_{4}(s,t)+2\mathcal{A}_{4}(t,u)+2\mathcal{A}_{4}(u,s)\nonumber \\
	&~=~ 2\int_{R_{s}}d^{2}z_{i}I(z_{i})+2\int_{R_{t}}d^{2}z_{i}I(z_{i})+2\int_{R_{u}}d^{2}z_{i}I(z_{i})\;,
	\end{align}
	with $R_s$, $R_t$ and $R_u$ disjoint, corresponding to the $s$, $t$ and $u$ channels in the field theory. On a tangential note, it is interesting to remark that cycling $(stu)$, which maps the individual integrals into each other, maps the saddle point $\hat{\tau}_2$ to points as depicted in figure \ref{fig:CongSubgroup}. Thus, if one is only interested in saddle points, one can choose a single integral in the above and integrate $\tau$ over the fundamental domain of the congruence subgroup $\Gamma_0(2)$, also depicted in figure \ref{fig:CongSubgroup}.   
	
	Using Feynman parameters $\alpha_i$ and excising vertex operator collisions, the plane wave factor in the $R_{s}$ region becomes
	\begin{equation}
	\prod_{i<j}\exp\big\{-k_{i}\cdot k_{j}G(z_{ij}|\tau_{2})\big\}~\longrightarrow~\exp\big\{-t(s_{12}\alpha_{1}\alpha_{3}+s_{14}\alpha_{2}\alpha_{4})\big\}\qquad\text{as }\alpha'\rightarrow0\;,\label{eq:FourPointFieldSChan}
	\end{equation}
	in the field theory limit. The other regions are obtained by cycling $(stu)$. In each region, the saddle point behaviour is destroyed. Indeed, whilst on the Feynman parameter $\alpha$-subspace there are still saddle points at $\alpha_{1}=\alpha_{3}=-\frac{t}{2u}$	and $\alpha_{2}=\alpha_{4}=-\frac{s}{2u}$ (which of course correspond to aforementioned Gross-Mende extremal points at $v_{1}=-\frac{t}{2u}$, $v_{2}=\frac{1}{2}$ and $v_{3}=\frac{1}{2}-\frac{t}{2u}$ and $v_{4}=0$), the Schwinger parameter $t$ has no extremum, as stated in the previous subsection.
	
\subsection{Softening in the Truncated Theory}
We claim that our truncated theory, in certain kinematical regimes, also has stringy saddle points.
As an example, let us take the theory truncated to $N=1$. For now we will consider the case with full $U(1)$ gauge symmetry, commenting on the case with discrete gauge symmetry later. By inserting four gauge invariant integrated vertex operators as discussed above, the $s$-channel amplitude of our mock string theory will contain the plane wave factor
\begin{align}
	\prod_{i<j}\exp\big\{-k_{i}\cdot k_{j}G(z_{ij} & =i\hat{v}_{ij}|\tau=A+iT)\big\}~=~\exp\bigg\{\frac{\alpha'}{2}\big(sv_{1}(v_{3}-v_{2})+tv(v_{2}-v_{1})(1-v_{3})\nonumber \\
	&\qquad\qquad+~\sum_{i<j}\sum_{k=0,1}s_{ij}\cos(2\pi(u_{ij}+k A v_{ij}))e^{-2\pi T |v_{ij}|}\;\big)\bigg\}\;.
	\end{align}
	Recall that the $u_i$ parameters are here associated to the gauge invariant vertex operators, and are to be integrated over $[0,1]$.
	
	The $v_i$-subspace contains the same saddle points $\hat{v}_i$ as in section \ref{sec:Gross-MendeSoftening}, so that
	\begin{align}
	\prod_{i<j}\exp\big\{-k_{i}\cdot k_{j}G(i\hat{v}_{ij}|A+iT)\big\}~&=~\exp\bigg\{\frac{\alpha'}{2}\big(-\pi\tau_{2}\frac{st}{2u}\nonumber \\
	&+\sum_{i<j}\sum_{k=0,1}s_{ij}\cos(2\pi(u_{ij}+k A \hat{v}_{ij}))e^{-2\pi T |\hat{v}_{ij}|}\;\big)\bigg\}\;.
	\end{align}
First note that even with $N=1$, we obtain an exponentially good approximation to the torus Green function at large $\tau_2$. Now, thanks to the exponential corrections which crucially come with varying signs, the above expression has saddles. To be explicit, let us work in the limit $t\rightarrow0$ and $u\sim-s$. Then at leading order, one finds a deep extremum at $u_{1}=u_{2}=1/2$, $u_{3}=0$ and $A=0$. At these values, the above becomes
	\begin{align}
	\prod_{i<j}\exp\big\{-k_{i}\cdot k_{j}G(i\hat{v}_{ij}|A+iT)\big\}~\simeq~\exp\bigg\{\frac{\alpha'}{2}\big(\pi T \frac{t}{2}+8se^{-\pi T}\big)\bigg]\;,
\end{align}
	so there is an extremum at $T=\hat{T}$ given by
	\begin{equation}
	\exp(-\pi\hat{T})=\frac{t}{16s}\;,
	\end{equation}
	matching the Gross-Mende result in \eqref{eq:GMapp}. 
	
	We conclude that the first string oscillations, and the corresponding leading exponential correction to the worldline Greens function, explain  string-like amplitude suppression in fixed-angle scattering.
	
	Interestingly this is even true if we also truncate to the  mock string theory with only discrete gauge symmetry $\mathbb{Z}_n$ Then, the integral over the $u_i$ becomes a sum over $u_i=0,1/n,\ldots,(n-1)/n$. Because the previous saddles for the $u_i$ occurred at $0$ and $1/2$, it is clear that the minimal $\mathbb{Z}_2$ discrete gauge symmetry is sufficient to maintain the same saddle point behaviour\footnote{Although this conclusion will change for higher point amplitudes where the saddle point behaviour changes.}.
	
	\subsection{Two Point Amplitude}
	Finally in this section let us briefly discuss the two-point one loop amplitude between the lightest scalar vertex operators. As a reference, in string theory this is
	\begin{equation}
		\mathcal{A}_2(s) ~=~ \int d^2_{\mathcal{F}} \tau \int_{T(\tau)} d^2 z \langle V(0) V(z) \rangle_{T(\tau)} ~=~ \int_{\mathcal{F}} d^2 \tau \int_{T(\tau)}d^2 z \mathcal{Z}(\tau)e^{\frac12 sG(z;\tau)}\;,
	\end{equation}
	where we define $s=-(k_1+k_2)^2$ for incoming momenta $k_1$ and $k_2$, $\mathcal{F}$ is the fundamental domain of $SL(2,\mathbb{Z})$ and $G(z;\tau)$ is given in \eqref{eq:GreensTheta}. This expression requires that the vertex operators have correct conformal weight, equivalent to insisting that the particles are on-shell (eg $k_1^2 = k_2^2 = -2$ in the case of tachyons). Furthermore, momentum conservation then implies
	$k_{1}+k_{2}=0$, which completely fixes $s$. Therefore, in order to be able to vary $s$, we will adopt the usual trick of breaking Lorentz invariance.
	
	Now consider the mock string theory with $N=1$ with continuous $U(1)$ gauge symmetry. It has one-loop two-point amplitude
	\begin{equation}
		\mathcal{A}_2(s)~=~ \int_0^\infty dT \int_{-1/2}^{1/2} dA \int_{[0,1]^2} dv du \mathcal{Z}(T,A) e^{\frac12 s (G_0^\circ(v,A,T)+e^{2\pi i u}G_1^\circ(v,A,T))}
	\end{equation}
	with $G_m^\circ$ as in \eqref{eqn:GmCirc}. Consider large $s$, in which we would like to use a saddle point approximation. The Green function $G_0^\circ$ contributes as in conventional field theory a saddle in the $v$-subspace at $\hat{v} = 1/2$ (this persists even when more $G_m$ are added due to symmetry). Again, whilst the Schwinger parameter $T$ of the field theory amplitude  does not have any extrema, here the addition of exponential terms to the Green function is sufficient for there to be a saddle even for $T$. Indeed, the saddle lies at ${A}={u} = 1/2$ and constant positive ${T}$. It therefore follows that this amplitude is exponentially suppressed at large $s$.
	
\section{Generalisations}
\label{sec5}
	In the previous sections, we considered worldline theories whose towers of particles corresponded to those of Kaluza-Klein vibrations of the string, and its truncations, and as such they had integer masses. However the interesting aspect of this approach is that there is no reason to remain confined to stringy geometry. One can contemplate extensions that contain towers of particles with different arrangements of masses. In this section we present some possibilities. Some of these inevitably will correspond to string theory in other guises, while for others the geometric interpretation is not obvious. We will proceed sequentially, first generalising to worldline theories in which the masses of the particle towers are not necessarily integers, and then discussing mass degeneracies in the towers. We then briefly consider  worldline fields whose KK indices span a lattice of more than one dimension. Such theories presumably capture some of the properties of higher dimensional objects. 	Finally we discuss worldline fermions, which would be required to produce  space-time fermions but can also be considered independently  \cite{Brink:1976uf,Strassler:1992zr}. 
	
	\paragraph{Generalised towers:} Our first examples consist of deforming the mass of the field $X_m$ to be some function $f_m=f(m)$ rather than $m$. This leads to a natural generalization of Eq.\eqref{eqn:Action}:
	\begin{equation} 
		S ~=~ \int d\tau \sqrt{g} \sum_{m=-N}^N \bigg( g^{\tau\tau} |D_\tau X_m|^2+ (2\pi)^2 f_m^2 |X_m|^2  \bigg) \bigg\}\qquad \text{ with }X_{-m}=X_m^\dagger\;.
	\end{equation}
	By choosing the field $X_0$ to remain massless (i.e. set $f_0 = 0$), it can still be interpreted as a coordinate field that embeds the worldline into a Poincar\'e-invariant spacetime. There is then great freedom to choose values for the remaining  masses $f_m$ such that at low energies the theory automatically reverts to the standard worldline theory.
	
	We can repeat the story outlined in section \ref{sec2} for these more general towers. Focussing on the partition function\footnote{Note that the theory's Green functions will also change relative to before.}, the calculations proceed exactly as before, but with $f_m$ in place of $m$. In particular, the partition function in one dimension now has the form
	\begin{equation} 
		\mathcal{Z}(A,T) ~=~ \frac{1}{T^{1/2}}\prod_{m=1}^N \frac{1}{|\sin(\pi f_m (A+iT))|^2} \;. 
	\end{equation}
	Taking the formal limit $N\rightarrow \infty$ will be of particular interest to us. As before, one should regularise this product. We illustrate this with two simple examples:
	\begin{enumerate}
		\item Let $f_m=m+\alpha$ for $m\geq 1$ and some $\alpha\geq 0$. Then the regularised partition function is 
		\[ \mathcal{Z}(\tau=A+iT)~=~ \frac{1}{2T^{1/2}}\bigg|q^{\frac14(\alpha^2 - \alpha +\frac16)} \prod_{m\geq 1} \big(1-e^{2\pi i m \alpha} q^m\big)\bigg|^{-2}\;, \]
		where $q=e^{2\pi i \tau}$ and the polynomial comes about from regularising using the Hurwitz zeta function $\zeta(s,\alpha)$, which has $\zeta(-1,\alpha)=-\frac12 B_2(\alpha)=-\frac12(\alpha^2-\alpha + \frac16)$.
		
		\item Let $f_m = \sqrt{m^2 + \alpha^2}$ for $m\geq 1$ and some $\alpha\geq 0$. Then the regularised partition function is
		\[ \mathcal{Z}(\tau=A+iT) ~=~ \frac{1}{2T^{1/2}} \bigg| q^{c_\alpha} \prod_{m\geq 1} \big(1-q^{\sqrt{m^2+\alpha^2}}\big) \bigg|^{-2}\;, \]
		with $c_{\alpha} = \frac{1}{(2\pi)^2}\sum_{k=1}^\infty \int_0^\infty dx e^{-k^2 x-\frac{\pi^2\alpha^2}{x}}$.
	\end{enumerate}
	The first of these examples corresponds in string theory to the string modes being shifted to $\alpha_{m+\alpha}$, and so is reminiscent of what happens in an orbifold theory. The second example is reminiscent of a string in a plane wave background, see \cite{Berg:2019jhh,Bergman:2002hv}.
	
	By combining different towers, one can begin to construct partition functions. For example, we can formally construct the half-integral Jacobi theta functions $\vartheta_1$ and $\vartheta_2$ by combining two worldline towers, with masses $f_m = (m-1+\alpha)/2$ and $f_m = (m-1-\alpha)/2$ for $m\geq 1$. This has associated partition function
	\begin{align}
		 \mathcal{Z}(\tau) &~=~ \frac{1}{2\sqrt{T}} \bigg| q^{-\zeta(-1,-\alpha)/2} \prod_{m=0}^\infty (1-q^{m-\alpha}) \bigg|^{-2} \cdot \frac{1}{2\sqrt{T}} \bigg| q^{-\zeta(-1,\alpha)/2} \prod_{m=0}^\infty (1-q^{m+\alpha}) \bigg|^{-2} \nonumber \\
		&~=~ \frac{1}{4T}  \bigg| \frac{\vartheta\begin{bsmallmatrix}\alpha-\frac12 \\ \frac12 \end{bsmallmatrix}}{\eta}\bigg|^{-2}\;.
		\end{align}
	It is not possible to construct the integral Jacobi theta functions $\vartheta_3$ and $\vartheta_4$ with these scalar Lagrangians - one would need anti-periodic boundary conditions, and towers of worldline fermions. However the story would be morally the same.
	\paragraph{Mass Degeneracies:} Expressing the partition function as a regularised product of $\sinh$ functions leads to a connection with the Bocherds' product formula \cite{Borcherds95}. To make this precise, let us first refine our notation to include a degeneracy of masses, so that the Lagrangian is (where the worldline $\tau$ here should not be confused with the Teichm\"uller parameter $\tau$)
	\begin{equation} 
		S \,= \, \frac12\int d\tau \sqrt{g} \bigg\{g^{\tau\tau} \sum_{k=1}^{c(0)}|D_\tau X_{0,k}|^2+2g^{\tau\tau} \sum_{m=1}^N \sum_{k=1}^{c(m)}\bigg( |D_\tau X_{m,k}|^2+ (2\pi)^2 f_{m}^2 |X_{m,k}|^2  \bigg) \bigg\}\;,
	\end{equation}
	where the fields $X_{m,1},\ldots,X_{m,c(m)}$ all have the same mass $f_{m}$. The associated partition function is now
	\begin{equation}
		\mathcal{Z}(\tau:=A+iT) ~=~ \frac{1}{T^{c(0)/2}}\bigg|q^{-C}\prod_{m=1}^\infty (1-q^{f_m})^{c(m)}\bigg|^{-2}\;,
	\end{equation}
	where $C$ is the regularised sum $\frac12\sum_{m=1}^\infty f_m c(m)$. We will assume that we can tune the constant $C$ by including a cosmological term $\int \sqrt{g} \Lambda$ in the action. Then $C$, $f_m$ and $c(m)$ essentially become free parameters. It is well known that there is a rich theory of expressing automorphic forms as such products. For example according to Bocherds in \cite{Borcherds95}, given a weakly holomorphic modular form $f(\tau)=\sum c(m)q^m$ of weight 1/2 for $\Gamma_0(4)$ whose integer coefficients vanish unless $n=0$ or $1$ mod $4$, the function
	\begin{equation}
		\Psi(\tau) ~=~ q^{-h}\prod_{m\geq 1}(1-q^m)^{c(m^2)}\;,
	\end{equation}
	is a meromorphic modular function for some character of $SL_2(\mathbb{Z})$ of integral weight. Here, $h$ is the constant term of $f(\tau) \sum H(n)q^n$ where $H(n)$ is the Hurwitz class function, with $H(0)=-1/12$. This correspondence between modular functions is an example of a so-called theta lift \cite{Fleig:2015vky}.  It gives us a way to express certain modular forms as infinite products, from which we can reverse-engineer a partition function with nice modular properties by choosing appropriate masses $f_m$ and degeneracies $c(m)$. (A limitation so far is that the $c(m)$ must necessarily be positive, but adding worldline fermions will allow this option.)
	
	In fact we have effectively encountered this already in section \ref{sec2}, because  the case  $f(\tau)=\vartheta_3(\tau)=1+2q+2q^4+2q^9+\cdots$, lifts to $\Psi = q^{1/12}\prod_{m\geq 1}(1-q^m)^2 = \eta(\tau)^2$. For a less trivial example, it is possible to lift a weight $1/2$ modular form to the Eisenstein series $E_6(\tau)$, which furnishes a product expansion of the form 
	\begin{equation}
		E_6(\tau) ~=~ (1-q)^{504}(1-q^2)^{143388}(1-q^3)^{51180024}\cdots
	\end{equation} 
	where the exponents are all known and positive. We refer to section 15 of \cite{Borcherds95} for more details. In this way, we can design a partition function which looks like
	\begin{equation}
		\mathcal{Z}(\tau)~=~\frac{1}{T^{6} |E_6(\tau)|^2}\;.
	\end{equation}

	\paragraph{Multiple indices:} Although formally covered by the previous cases, it is interesting to rearrange our notation in order to allow for worldline theories whose indices span a lattice. For example consider the following action
	\begin{equation}
		 S[g,X_m] = \frac12\int d\tau \sqrt{g} \bigg\{g^{\tau\tau} (\partial_\tau X_0)^2+2\sum_{(m,n)\neq (0,0)}^N \bigg( g^{\tau\tau}|D_\tau X_{m,n}|^2 + (2\pi)^2 (m^2+n^2) |X_{m,n}|^2  \bigg) \bigg\}\;,
	\end{equation}
	(supplemented with $X_{-m,-n}=X_{m,n}^\dagger$) where now the worldline fields have two indices $X_{m,n}$ which span the lattice $\mathbb{Z}^2$. Such an action is achieved for example by considering a object with topology $S^1 \times S^1$. In this case, the  ``closed bucatini'' partition function is
	\begin{equation}
		\mathcal{Z}(\tau)~=~\frac{1}{T^{1/2}} q^{-E^*(i,-1)}\prod_{(m,n)\neq (0,0)}(1-q^{\sqrt{m^2+n^2}})\;,
	\end{equation}
	where $E^*(\tau,s)=\frac12 \sum_{(c,d)\neq (0,0)}\frac{\tau_2^s}{|c\tau+d|^{2s}}$ is an Eisenstein series, to be evaluated at $\tau=i$.
	
		\paragraph{Extension to Fermions:}
	A natural generalisation is to add fermions into the worldline theory. If one is interested in describing spacetime fermions, then the 
	required modification will turn the worldline theory into a supersymmetric one \cite{Brink:1976uf,Strassler:1992zr,Bastianelli:2007pv}. However worldline supersymmetry is not mandatory: that is 
fermions can exists in a worldline theory without forming part of some supermultiplet (and indeed the same is true of worldsheet fermions in heterotic string theory as well).  
Nevertheless we will begin by writing the superpartner for the worldline action.  
	That is, following from \eqref{eqn:Action}, let $\psi_m$ be a tower of two-component Grassmann fields, and let us consider propagation on a line with Teichm\"uller parameter $T$ and the following action:
	\begin{equation}
	S_\psi ~=~ \frac{i}{8\pi \ell_s^2} \sum_{m\in \mathbb{Z}+\nu} \int_0^1 d\tau \,\psi_m^\dagger \begin{pmatrix} D+2\pi m T & 0 \\ 0 & D-2\pi mT \end{pmatrix}\psi_m\;,
	\end{equation}
	where $D\psi_m=(\partial_\tau - 2\pi i m T)\psi_m$ is the same covariant derivative as before, and the above is to be supplemented with $\psi_{-m}=\psi_m^\dagger$. Due to the latter condition, we must restrict the parameter $\nu$ to either $\nu=0$ or $\nu=1/2$. Not surprisingly from a string perspective, the above action would arise naturally from dimensionally reducing the kinetic term of a worldsheet fermion in which case the choices $\nu=0$ or $1/2$ correspond to an R or NS fermion.
	
	Following the same route as in section \ref{sec2}, the kernel corresponding to $\psi_m$ propagating on the line segment can easily be calculated. 
	Actually, since the on-shell action vanishes, this kernel is also essentially equal to the one-loop partition function. 
	We let $\mu=0$ or $1/2$ depending on whether $\psi_m$ is to be a Grassmann boson or fermion respectively. The final result is\footnote{Note that we are consistently using $\tau$ to be $A+iT$ whenever we refer to the partition function.}
	\begin{equation}
		K_m(\tau\coloneqq A+iT) ~=~ |\mu-\tau m|^2 \prod_{\substack{n \in \mathbb{Z} \\ n\neq 0 }} \big| 1 - \frac{\tau m}{n+\mu} \big|^2 ~=~ \bigg| \frac{\mu\sin(\pi (m\tau-\mu))}{\sin(\pi \mu)} \bigg|^2\;,
	\end{equation}	
	up to constant factors. In the string theory, the entire tower is present (with all possibilities of $\mu,\nu\in \{0,1\}$ being required to build a modular invariant ${\cal Z}$), and again using zeta function regularisation results in an effective kernel for the tower being
	\begin{equation}
	\begin{aligned}
	&K^{\mu,\nu}(\tau) = \bigg|\frac{\vartheta\begin{bsmallmatrix} \mu \\ \nu \end{bsmallmatrix}(\tau)}{\eta(\tau)}\bigg|^2\;.
	\end{aligned}
	\end{equation}
	Of course, this is entirely standard. But note that we can identify the zero-mode, $\psi_0$, in the $(\mu,\nu)=(0,0)$ sector, as being responsible for the vanishing of $\vartheta_1$ in string theory.
	
	We can now adjust the fermion tower in a way that is not obviously string-like to access a wider range of partition functions. In particular, it is now possible to extend our discussion of the Borcherds product formula to the case of alternating signs for the exponents $c(m)$ by tuning the bosonic and fermionic tower respectively. For example we can use the $(\mu,\nu)=(0,0)$ version of the above and omit the $m=0$ term in the $(\mu,\nu)=(0,0)$ sector which is responsible for the vanishing of that tower's partition function. Upon doing this, we can use the Bocherds product formula in more generality to find more partition functions. For example, we can now produce the modular invariant partition function
	\begin{equation}
	\frac{1}{T^4 |E_4(\tau)|^2}\;,
	\end{equation}
	where the Eisenstein series $E_4(\tau)$ has a product expansion with even and odd powers of $(1-q^n)$ \cite{Borcherds95}.
	
\section{Conclusions}
\label{conclusions}

Worldline theories provide an alternative formalism for perturbative quantum field theory. We have shown that adding a tower to the worldline with a spectrum of \textit{worldline} masses  leads to a non-local theory in spacetime with special properties. Infinite towers have a close relation to string theory, and the presence of such a tower may signal that the theory is not truly a theory of particles, but rather consists of higher dimensional objects that have been dimensionally reduced onto the worldline. Indeed, if the masses of the particle tower are integrally spaced and one-dimensional, we can interpret the fields of the tower to be the harmonics of a vibrating string. This worldline description then provides a setting to directly compare local QFT with string theory. Moreover the spectrum of the worldline states can be seen to be the crucial factor in making the theory finite. As soon as the worldline action is made compact, \eqref{eq:Sm}  has explicit $T\rightarrow 1/T$ symmetry, so that finiteness can be considered to be ``baked in'' to the theory already at this stage, becoming manifest in loop diagrams. 
	
	Among the other special properties that we wish to highlight is the behaviour of the theory's amplitudes. We have shown that even when the tower is severely truncated to 
a particle theory that has only a single internal mode, the four point loop amplitude of the theory behaves identically (that is its fixed-angle scattering has a saddle point and exponential suppression) to a string in the Gross-Mende hard scattering limit. Such behaviour is hard to come by, being impossible among unitary local quantum field theories. Actually, we can very simply illustrate the difficulty of achieving a saddle point behaviour by the following. Imagine some possibly non-local quantum field theory which contains a spacetime propagator which can be written as $\pi(p^2)=\int_0^\infty dt \exp(-p^2 F(t)-m^2 t)$, such that $\pi(p^2)$  reduces to the Feynman propagator at small $p^2$ \cite{Abel:2019ufz,Abel:2019zou}. A nontrivial example is $\pi(p^2)=e^{-p^2}/(p^2+m^2)$ \cite{Siegel:2003vt,Biswas:2005qr,Biswas:2011ar,Buoninfante:2018mre} which corresponds to $F(t)=t+1$. More generally this defines a wide class of possible non-local field theories, and yet, leaving aside the strong constraints on $F(t)$ coming from unitarity and issues concerning Wick rotation, it is straightforward to show that there can be no saddle point in the corresponding four point loop amplitude. Achieving saddle point behaviour really does require something more delicate than simply modifying a particle propagator in this way. Our theory provides such behaviour.
	
	On the other hand there seems to be potential to treat other kinds of objects using this formalism, and the approach provides a simple framework for developing theories with useful arithmetical properties, whose geometry may be less clear. In this sense the possibilities of the worldline formalism have not yet been fully explored, as there  
is clearly scope to consider all kinds of internal SHO degrees of freedom carried by the propagating particle. 
	As we have noted, there is for example no need to assume integrally spaced worldline masses in the tower. Indeed, if we consider such worldline masses to be arbitrary (up to the requirement that the theory reproduces a local worldline theory at low energies), there are many interesting possibilities, just a few of which have been mentioned in this paper. This generalised framework allows us to pass from theories that are geometrically motivated to ones that do not admit an obvious geometric origin such as a worldsheet (at least a priori). Conversely, one can reverse-engineer theories (by studying some appropriate tower) that do not have obvious geometrical interpretation. One noteable class of theories of this kind  we presented is based on a Borcherds product procedure. These also appear to enable ``baked in'' finiteness, in that their partition functions consist of modular functions. But their only obvious interpretation is that of a particle with a particular infinite tower of internal degrees of freedom: their geometrical meaning is unclear.
	
	Developing a full programme of perturbation theory  in such a construction remains a challenge, because we have not developed a general approach for interactions, other than introducing effective vertex operators for the tower. This allows for external emission, and hence is sufficient for one-loop diagrams. The question of matching propagators at internal vertices is less obvious. In principle, one could form worldgraphs and find the Green functions for the entire graph \cite{Dai:2006vj}. But taking inspiration from string theory, it is not clear that this is the correct approach. For example, in closed string splitting the centre of mass is the coordinate field $X_0$ which defines the worldline. But then as the string splits, the worldline must split into two worldlines, discontinuously and not at a single vertex. Reminiscent of problems that are encountered in string field theory, it is then a delicate question how to match propagators at a vertex. This will be addressed in future work. 
	
\subsection*{Acknowledgements} We are extremely grateful to Nicola Dondi for conversations and collaboration. We also wish to thank James Edwards for helpful comments. DL is supported by an STFC studentship.

\end{document}